\documentclass[aps,twocolumn,showpacs,showkeys,amsmath,amssymb,nofootinbib,superscriptaddress,prc,floatfix]{revtex4-1}
\usepackage{graphicx}
\usepackage{subfigure}
\usepackage{bm}

\makeatletter
\newcommand\erfc{\mathop{\operator@font erfc}\nolimits}
\def\slashchar#1{\setbox0=\hbox{$#1$}
   \dimen0=\wd0 \setbox1=\hbox{/} \dimen1=\wd1
   \ifdim\dimen0>\dimen1 \rlap{\hbox to \dimen0{\hfil/\hfil}} #1
   \else  \rlap{\hbox to \dimen1{\hfil$#1$\hfil}} / \fi}

\begin{document}
\title{Interferometry radii in heavy-ion collisions at $\sqrt{s}=200$GeV and $2.76$TeV}
\author{Piotr Bo\.zek}
\email{Piotr.Bozek@ifj.edu.pl}
\affiliation{The H. Niewodnicza\'nski Institute of Nuclear Physics,
PL-31342 Krak\'ow, Poland} 
\affiliation{
Institute of Physics, Rzesz\'ow University, 
PL-35959 Rzesz\'ow, Poland}
\date{\today}

\begin{abstract}
The expansion of the fireball created in 
 Au-Au collisions at $\sqrt{s}=200$GeV and 
Pb-Pb collisions at $2.76$TeV 
 is modelled using the  relativistic viscous hydrodynamics. The
 experimentally  observed interferometry radii are well reproduced. 
Additional pre-equilibrium flow improves slightly the results for the lower 
energies studied.
\end{abstract}



\maketitle

\section{Introduction}

The matter created in ultrarelativistic heavy-ion collisions undergoes a 
rapid expansion. During the expansion substantial transverse flow is generated
modifying the transverse momentum spectra of emitted particles. 
The measured  spectra and elliptic flow at
the BNL Relativistic Heavy Ion Collider 
\cite{Arsene:2004fa,*Back:2004je,*Adams:2005dq,*Adcox:2004mh} can be 
  described within relativistic hydrodynamic models \cite{Kolb:2003dz,*Huovinen:2006jp,*Hirano:2008aj,*Ollitrault:2010tn,Florkowski:2010zz}.
 The hydrodynamic behavior of the bulk of the matter created in 
  Pb-Pb collisions at $\sqrt{s}=2.76$TeV is confirmed  in recent experiments 
at the CERN
 Large Hadron Collider (LHC) \cite{Aamodt:2010pa,alice:2010mr}.
The magnitude of the observed elliptic flow indicates that  the dense
matter behaves as an almost 
 perfect, thermalized 
 fluid, with small shear 
viscosity \cite{Romatschke:2007mq,*Song:2008hj}.

The size and the life-time of the source can be estimated from the 
Hanbury Brown-Twiss  (HBT) radii extracted from Bose-Einstein correlations of identical particles \cite{Bertsch:1988db,*Pratt:1986cc,*Wiedemann:1999qn,*Heinz:1999rw,*Lisa:2005dd}. The description of the HBT radii in dynamical 
models has been the subject of intensive studies. Many calculations lead 
to an overestimate of the ratio of  $R_{out}$ to $R_{side}$ radii, 
which was
 termed as the HBT puzzle. 
 Capturing correctly the values of the $R_{side}$ and $R_{long}$ radii in a model calculation  amounts to a satisfactory description of the 
 the size and the 
life-time of the system. The radius $R_{out}$ 
(and the ratio $R_{out}/R_{side}$) 
are more sensitive to the collective flow of the matter at the freeze-out. 
The same is true for the dependence of the radii on the momentum of 
the pion pair. Experimental observations yield $R_{out}/R_{side}\simeq 1$, 
contrary to expectations for a system undergoing a first order phase
 transition \cite{Rischke:1996em}. Reversing this argument, the experimental
data indicate that the equation of state of the hot matter 
is hard, without a soft point
 \cite{Broniowski:2008vp,*Kisiel:2008ws,Pratt:2008qv,*Pratt:2008sz}. Other effects that 
influence the ratio $R_{out}/R_{side}$ are the  shear viscosity 
\cite{Pratt:2008qv,*Pratt:2008sz,Bozek:2009dw}, the partial chemical equilibrium
\cite{Hirano:2002ds}, the early start of the hydrodynamic expansion 
($0.1$-$0.25$fm/c)
\cite{Broniowski:2008vp,Pratt:2008qv,*Pratt:2008sz} or the presence of some 
pre-equilibrium flow \cite{Karpenko:2009wf}.


We simulate central heavy-ion collisions at the top RHIC and LHC energies 
using
relativistic viscous hydrodynamics, with or without pre-equilibrium flow.
 We calculate the transverse momentum spectra and HBT radii. We obtain a
 satisfactory agreement with the experiments, the additional initial flow 
improves the  $R_{out}/R_{side}$ ratio at RHIC energies but 
has a smaller effect for Pb-Pb collisions at the LHC.

\section{Hydrodynamic model}

We use the second order relativistic viscous hydrodynamics \cite{IS} to model
 the expansion of fireball.
In relativistic viscous hydrodynamics the energy-momentum tensor is
\begin{equation}
T^{\mu\nu}=(\epsilon+p)u^\mu u^\nu- p g^{\mu \nu} +\pi^{\mu \nu}+ \Pi 
\Delta^{\mu \nu}  \ ,
\label{eq:tmunu}
\end{equation}
 with stress corrections from shear
$\pi^{\mu\nu}$ and bulk $\Pi \Delta^{\mu \nu}$ viscosities,
 $\Delta^{\mu \nu}= g^{\mu\nu}-u^\mu u^\nu$. The stress corrections are the
solutions of the second order relativistic viscous 
hydrodynamic equations \cite{IS,*Teaney:2003kp,*Muronga:2001zk,*Song:2007ux,*Baier:2006gy,*Dusling:2009df,*Chaudhuri:2006jd,*Dusling:2007gi,*Romatschke:2009im,*Teaney:2009qa,*Chaudhuri:2006jd,*Luzum:2008cw,Romatschke:2007mq,Bozek:2009dw,Schenke:2010rr}
\begin{equation}
\Delta^{\mu \alpha} \Delta^{\nu \beta} u^\gamma \partial_\gamma \pi_{\alpha\beta}=\frac{2\eta \sigma^{\mu\nu}-\pi^{\mu\nu}}{\tau_{\pi}}-\frac{1}{2}\pi^{\mu\nu}\frac{\eta T}{\tau_\pi}\partial_\alpha\left(\frac{\tau_\pi u^\alpha}{\eta T}\right) 
\end{equation}
and
\begin{equation}
 u^\gamma \partial_\gamma \Pi=\frac{-\zeta \partial_\gamma u^\gamma-\Pi}{\tau_{\Pi}}-\frac{1}{2}\Pi\frac{\zeta T}{\tau_\Pi}\partial_\alpha\left(\frac{\tau_\Pi u^\alpha}{\zeta T}\right)  \ , \end{equation}
with  \begin{equation}
\sigma_{\alpha\beta}=\frac{1}{2}\left( \nabla_\alpha  u_\beta
+\nabla_\beta u_\alpha -\frac{2}{3}\Delta_{\alpha \beta}\partial_\mu u^\mu\right)\ ,
\end{equation} 
$\eta$ and $\zeta$  are the shear and bulk viscosity coefficients and two 
relaxation times  $\tau_\pi $ and $\tau_\Pi$ appear.
We use the Navier-Stokes initial conditions for the stress tensor 
$2 \pi^{xx}(\tau_0)=2 \pi^{yy}(\tau_0)=-\pi^{zz}(\tau_0)= 4 \eta/3 
\tau_0$  corresponding to the longitudinal Bjorken flow and $\Pi(\tau_0)=0$.

The initial profile of the entropy density at the impact parameter $b$ 
is obtained from the Glauber model
\begin{equation}
s(x,y,b)=s_0 \frac{ (1-\alpha)\rho_{WN}(x,y,b)+2 \alpha 
\rho_{bin}(x,y,b) }{ (1-\alpha)\rho_{WN}(0,0,0)+2 \alpha 
\rho_{bin}(0,0,0) } \ , 
\label{eq:gm}
\end{equation}
where $\rho_{WN}$ and $\rho_{B}$ are the densities of wounded nucleons and  
 binary collisions respectively.
 The optical Glauber 
model densities are obtained with Wood-Saxon densities for the colliding
 nuclei
 $\rho_{WS}(r)= \rho_0/\left(\exp\left((r-R_a)/a\right)+1\right)$
 ($\rho_0=0.169$fm$^{-3}$, $R_a=6.38$fm, $a=0.535$fm, $b=2.1$fm for Au and $\rho_0=0.169$fm$^{-3}$, $R_a=6.48$fm, $a=0.535$fm, $b=2.2$fm for Pb) and the nucleon-nucleon
cross sections
 are $42$mb and $62$mb at $\sqrt{s}=200$GeV and $2.76$TeV 
respectively. The contribution of binary collisions is $\alpha=0.145$ 
at RHIC energies \cite{Back:2004dy}. The centrality dependence 
of the charged particle multiplicity in Pb-Pb collisions at the LHC 
\cite{Collaboration:2010cz} is used to fix
$\alpha=0.15$ at the higher energy. Such 
a  value of the parameter $\alpha$ leads to an apparent
overprediction of the multiplicity in central Pb-Pb collisions 
\cite{Bozek:2010wt,Aamodt:2010pb}, but
this is due to the incorrect value of the reference 
multiplicity in proton-proton collisions used in Ref. \cite{Bozek:2010wt}.
We use a realistic equation of state interpolating between
 the lattice QCD data at high temperatures 
 and an equation of state of a gas of 
hadrons at lower temperatures \cite{Chojnacki:2007jc}. 
The 
use of an equation of state without a soft point is essential 
in reproducing the
femtoscopy data \cite{Broniowski:2008vp,Pratt:2008qv}.
 
The initial time for the hydrodynamic evolution is $\tau_0=0.6$fm/c.
The  corrections to the pressure from shear viscosity $2 \eta / 3 \tau_0$
 to the transverse pressure at the initial time are of the order of $20$\%, 
hence the application of viscous hydrodynamics is justified. 
On the other hand, the 
starting time of the hydrodynamic evolution should be  defined by the 
early thermalization mechanisms that are not well understood.
Moreover,  some early flow can be generated 
 before the start of the  hydrodynamic evolution. 
The pre-equilibrium transverse flow in hydrodynamic models has been discussed 
by several authors
 \cite{Kolb:2002ve,*Chojnacki:2004ec,*Gyulassy:2007zz,Karpenko:2009wf,Vredevoogd:2008id}. The mechanism generating the initial 
flow is not known, and   the amount of the pre-equilibrium flow is 
a parameter of the calculation. Ref. \cite{Vredevoogd:2008id} gives 
a general result for the transverse flow generated  in the initial 
stage of the reaction, which could serve as an upper bound for the 
pre-equilibrium flow generated in the  expansion at 
the early stage with boost invariance, starting at zero time. 

To estimate the effects of the 
collective evolution of the system before $\tau_0$ and of the choice of the value of $\tau_0$,
 we compare two different calculation. The first scenario
 assumes that the evolution starts at $\tau_0=0.6$fm/c with zero initial flow
and in  the second scenario the initial flow is taken in the form of the 
universal flow for the early transverse acceleration \cite{Vredevoogd:2008id}.
For a traceless energy-momentum tensor  the components of the energy momentum 
tensor take a universal form
\begin{equation}
\frac{T^{0i}}{T^{00}}=-\frac{\partial_i T^{00}}{2 T^{00}}  \tau_0
\label{eq:inifl}
\end{equation}
at the time $\tau_0$.

Starting from $\tau_0$ the evolution is governed by the relativistic
 viscous hydrodynamics with a  realistic equation of state. The 
initial energy density and pressure are given by the Glauber 
 profile of the entropy density
with corrections  from  shear viscosity given by the Navier-Stokes value used as
initial conditions for the second order Israel-Steward equations. 
The 
velocity profile at $\tau_0$ is matched to reproduce the ratio
 ${T^{0i}}/{T^{00}}$ predicted from the universality argument (\ref{eq:inifl}).
The effects of switching to a realistic equation of state and of the 
shear viscosity corrections after  $\tau_0$ on  the velocity field are 
small in the core of the fireball as compared to using a perfect fluid  with the 
ultrarelativistic gas equation of state.

\section{Results}

\begin{figure}
\includegraphics[width=.48\textwidth]{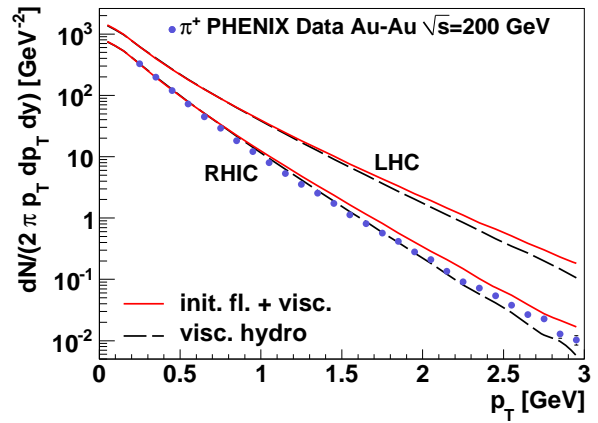}
\caption{(Color online) Transverse momentum spectra of $\pi^+$ in central collisions. The upper lines are for Pb-Pb collisions at $\sqrt{s}=2.76$TeV 
and the lower lines for Au-Au collisions at $200$GeV. The dashed lines are 
for standard initial conditions and  the solid lines are for the evolution
 starting with the  pre-equilibrium flow,
 all calculations in the relativistic viscous hydrodynamics; 
data for Au-Au collisions  are from the 
PHENIX Collaboration \cite{Adler:2003cb}.}
\label{fig:sp}
\end{figure}

\begin{figure}
\includegraphics[width=.48\textwidth]{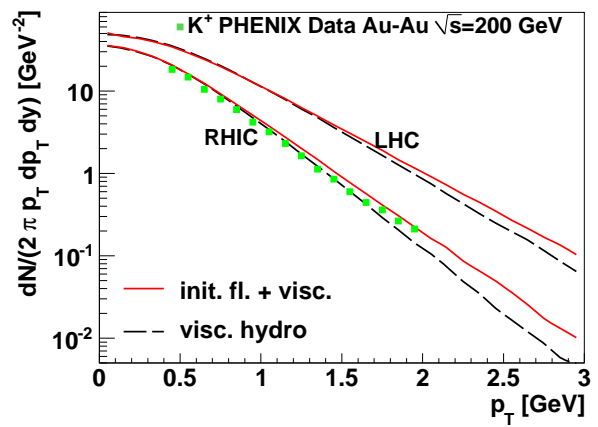}
\caption{(Color online)Same as Fig. \ref{fig:sp} but for $K^{+}$. }
\label{fig:spK}
\end{figure}

We set the ratio of the viscosity coefficients to the entropy to
 $\eta/s=1/4\pi$ for the shear viscosity and $\zeta/s=0.04$ for the 
bulk viscosity,  $\tau_\pi=\tau_\Pi=\frac{3\eta}{T s}$.
 The bulk viscosity is present only in the hadronic phase
 \cite{Bozek:2009dw}. The statistical emission and resonance decays are 
performed using the THERMINATOR program \cite{Kisiel:2005hn},
 at the freeze-out temperatures of $140$MeV  and $150$MeV for the calculations
without and with 
with pre-equilibrium flow respectively. 
The freeze-out with 
finite shear and bulk viscosities in the hadronic phase captures the main
characteristics of the dissipative hadronic rescattering phase at the end
of the expansion \cite{Bozek:2009dw}.
 The entropy density at the center of the
 fireball $s_0$ corresponds to a temperature of $400$ and $520$MeV 
at RHIC and LHC energies respectively.

\begin{figure}
\includegraphics[width=.4\textwidth]{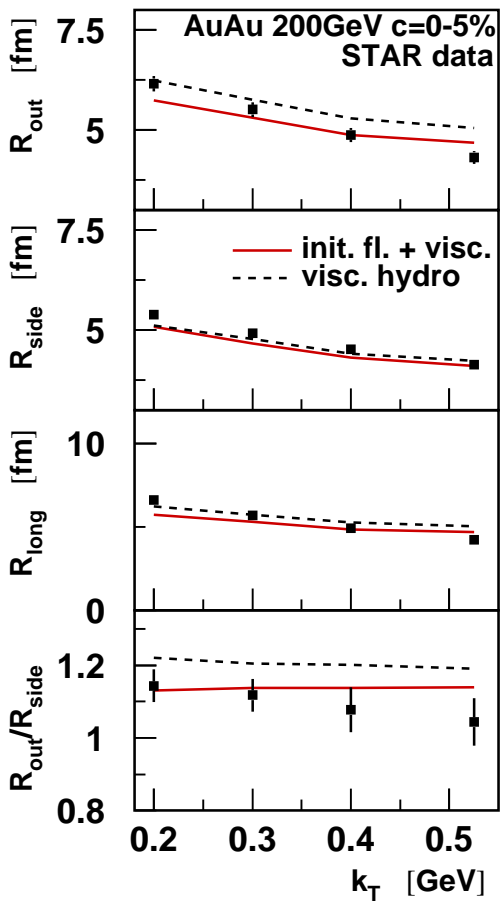}
\caption{(Color online) 
 STAR Collaboration data \cite{Adams:2004yc} (squares)  for the HBT radii 
in central  Au-Au collisions  at $\sqrt{s}=200$GeV
 compared to the results   from viscous 
hydrodynamics with standard (dashed lines) and pre-equilibrium flow
initial conditions (solid lines). }
\label{fig:hbt}
\end{figure}

In Figs. \ref{fig:sp} and \ref{fig:spK} are shown the transverse 
momentum spectra of pions and kaons. To compensate for the additional 
transverse flow, the freeze-out  temperature is increased to $150$MeV for 
calculations including the initial flow.  The introduction 
of  the initial flow leads to some hardening of the spectra as compared to
 the standard initial conditions.
 However, both calculations are close to
 the experimental data. 
The transverse momentum spectra at LHC energies are harder and the 
relative importance of the pre-equilibrium flow is smaller than at RHIC. 
The systems lives longer and most of the transverse flow is generated in the
hydrodynamic expansion of an almost perfect fluid.

\begin{figure}
\includegraphics[width=.4\textwidth]{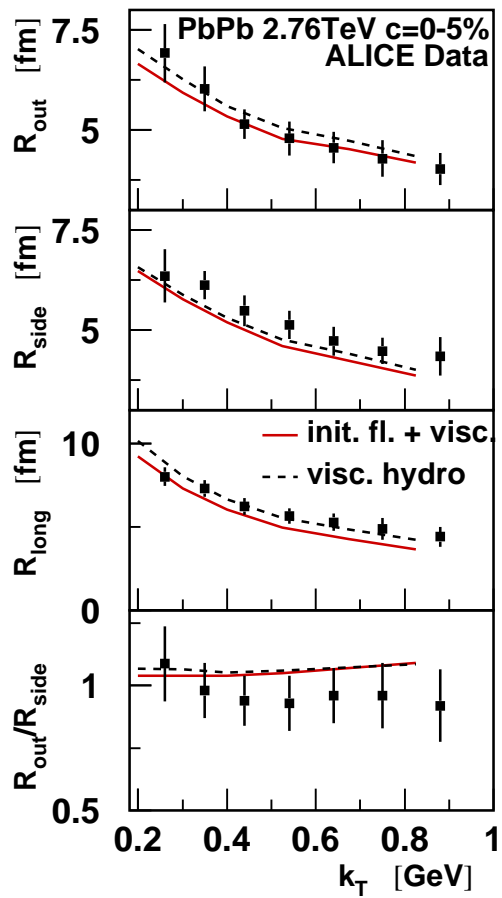}
\caption{(Color online) 
 ALICE Collaboration data \cite{alice:2010mr} (squares)  for the HBT radii 
in central  Pb-Pb collisions  at $\sqrt{s}=2.76$TeV
 compared to the results  from viscous 
hydrodynamics with standard (dashed lines) and pre-equilibrium flow  
initial conditions (solid lines). }
\label{fig:hbtign}
\end{figure}

For  events generated in THERMINATOR the  correlation function is 
constructed from same-event and mixed-event  pion pairs
\cite{Kisiel:2006is,*Kisiel:2006yv}. The three dimensional momentum 
correlation functions are fitted using a Gaussian parameterization and the three
HBT radii are extracted. The dependences of the three radii 
and of the ratio $R_{out}/R_{side}$ on the momentum of the pion
 pair are shown in Fig. \ref{fig:hbt} together with the experimental 
results of the STAR Collaboration \cite{Adams:2004yc}. The HBT radii are
 reproduced  in the viscous hydrodynamic calculation with standard initial
 conditions to within $8$-$15$\%, depending on the pair momentum. The additional
pre-equilibrium flow  (solid lines) brings the 
results closer to the data, reducing the discrepancy. The 
satisfactory description of the ratio $R_{out}/R_{side}$ shows that the 
collective flow generated in the model is realistic. The  values of
 the  three radii are close to the data, which 
indicates that the size and the life-time of the source
 are well described in the model.

The source created in Pb-Pb collisions at the LHC is larger and shows a 
stronger collective flow (Fig. \ref{fig:hbtign}). The viscous hydrodynamic
calculation reproduces the measured values of the radii
 to within $14$\%. The inclusion of the initial flow at LHC energies
 does not change 
 the agreement with the data. The earlier freeze-out for the calculation 
with initial flow leads to a reduction of the values of the radii, but 
the ratio $R_{out}/R_{side}$ is not modified significantly.
 We notice that the expansion at the LHC is dominated by the 
hydrodynamic stage, with a relatively smaller contribution from the 
pre-equilibrium phase. The strong flow observed in the femtoscopy data is 
quantitatively well 
described by the model calculation. The size of the fireball
at the freeze-out increases with the increasing multiplicity, as seen
 in the data.
In the simulation the freeze-out happens at $\tau\simeq 12$-$13$fm/c.

\section{Conclusions}

We make hydrodynamic calculations for central heavy-ion collisions
at   $\sqrt{s}=200$GeV (RHIC) and  $2.76$TeV (LHC). 
The fireball expansion is
 modelled using relativistic viscous hydrodynamics with the shear viscosity
 coefficient $\eta/s=1/4\pi$, and,  in the hadronic phase,
  the bulk viscosity of 
$\zeta/s=0.04$ is added. Calculations starting without initial flow 
at $\tau_0=0.6$fm/c 
describe the transverse pion  and kaon spectra at RHIC energies. The
 HBT radii at both colliders are reproduced to within $15$\%. The presence
 of an additional transverse flow in the initial state of the hydrodynamic
 expansion improves the description of the HBT data at RHIC energies. 
The deviations of the simulations from the data  are 
 of the order of the systematic errors quoted by the 
experimental Collaborations.

The paper presents the first quantitative description of the recently 
released ALICE Collaboration data on the HBT interferometry in Pb-Pb 
collisions at the highest available energy  \cite{alice:2010mr}. The 
relativistic viscous hydrodynamics gives a satisfactory agreement 
with the measurements. The effect of the  pre-equilibrium flow on the
 $R_{out}/R_{side}$ ratio is small.

\begin{acknowledgments}

The author  is grateful to Wojtek Florkowski for numerous discussions
on the HBT puzzle. 
The work is supported  by the
Polish Ministry of Science and Higher Education 
grant No.  N N202 263438.

\end{acknowledgments}

\bibliography{../hydr}

\end{document}